\documentclass[
aps,
prb,
10pt,
superscriptaddress,
floatfix,
twocolumn, 
notitlepage,
]{revtex4-2}
\usepackage{graphicx,graphics}
\usepackage{amsmath,amssymb,amsfonts}
\usepackage{latexsym,verbatim}
\usepackage{bm}
\usepackage{color}

\usepackage[percent]{overpic}
\usepackage[
breaklinks=true,
colorlinks,
citecolor=blue,
linkcolor=blue,
urlcolor=blue
]{hyperref}
\usepackage{orcidlink}
\usepackage[normalem]{ulem}
\usepackage[dvipsnames]{xcolor}
\usepackage{siunitx}
\newcommand{\Add}[1]{#1}
\newcommand{\Rem}[1]{}
\newcommand{\lb}{{\ell_{\rm B}}}
\newcommand{\rey}{{\rm Re}}
\newcommand{\reyi}{{\rm Re}^\infty}
\newcommand{\reyc}{{\rm Re_{\rm c}}}
\newcommand{\reycz}{{\rm Re}_{\rm c}^{(0)}}
\newcommand{\reyci}{{\rm Re}_{\rm c}^\infty}

\newcommand{\lbr}{{\tilde{\ell}_{\rm B}}}
\newcommand{\kc}{{k_{\rm c}}}
\newcommand{\fc}{{f_{\rm c}}}
\newcommand{\y}{\tilde{y}}
\newcommand{\lee}{\ell_{\rm ee}}
\newcommand{\vf}{v_{\rm F}}
\newcommand{\reycznum}{7696}
\newcommand{\reycinum}{48000}
\newcommand{\ainterp}{9.5}
\newcommand{\Wmin}{2}
\newcommand{\jmin}{6795}
\newcommand{\kcmin}{1.9}
\newcommand{\fmin}{0.68}
\begin{document}
\author{Marine Bastida \orcidlink{0009-0009-0390-3217}}
\affiliation{Département de Physique, École Normale Supérieure Paris-Saclay, 
91190 Gif-sur-Yvette, France}
\author{Álvaro Meseguer \orcidlink{0000-0002-2022-2001}}
\affiliation{Departament de Física, Universitat Politècnica de Catalunya - BarcelonaTech (UPC), 
Campus Nord B4-B5, C. Jordi Girona 1-3, 
08034 Barcelona, Spain}
\author{Iacopo Torre \orcidlink{0000-0001-6515-181X}}
\affiliation{Departament de Física, Universitat Politècnica de Catalunya - BarcelonaTech (UPC), 
Campus Nord B4-B5, C. Jordi Girona 1-3, 
08034 Barcelona, Spain}
\email{iacopo.torre@upc.edu}

\title{The \Add{linear} stability of electronic Poiseuille flow in two-dimensional materials}
\begin{abstract}
Motivated by the experimental observation of electronic Poiseuille flow in graphene [J.A. Sulpizio et al. Nature {\bf 576}, 75 (2019)] we analyze the linear stability of plane-Poiseuille flow of electrons in a two-dimensional material using a modified Orr-Sommerfeld equation.
We calculate the critical current needed to make the flow unstable as a function of the experimental parameters and characterize the most favorable situation, that is the one needing the lowest current density, to observe flow instability.
We predict the streamwise wavenumber and frequency at which the instability occurs and discuss the difficulties of an experiment aimed at probing this phenomenon. 
\end{abstract}
\maketitle
\section{Introduction}
\label{sect:intro}
The most elementary picture of electric current in a conductor is that of a fluid of electrons flowing inside the material. 
This picture turns out to be rigorous in very high quality materials inside a specific window of parameters \cite{torre_nonlocal_2015, levitov_electron_2016, bandurin_negative_2016, ho_theoretical_2018}: when interactions of electrons with defects and phonons of the lattice are weak and interactions among electrons are strong, the flow of current can be described using the equations of fluid mechanics \cite{torre_nonlocal_2015, narozhny_hydrodynamics_2015}.
More formally, this happens when the electron-electron scattering length $\lee$ is much smaller than any other length scale of the problem, i.e. much smaller than the momentum relaxation length $\ell_p$ and of any geometric scale $w$ of the sample.

Evidences of electron hydrodynamic transport have been reported in high-quality graphene \cite{bandurin_negative_2016, crossno_observation_2016, berdyugin_measuring_2019}, $\rm{GaAs/AlGaAs}$ quantum wells \cite{de_jong_hydrodynamic_1995,braem_scanning_2018}, ${\rm PdCoO_2}$ \cite{moll_evidence_2016}, and ${\rm WTe_2}$ \cite{aharon-steinberg_direct_2022}. 
\Add{The most important theoretical and experimental findings in electron hydrodynamics are summarized in recent review articles \cite{lucas_hydrodynamics_2018,fritz_hydrodynamic_2024,hui_hydrodynamics_2025}.}

The plane Poiseuille flow \cite{landau_fluid_1987} is probably the simplest, non-trivial, solution of the Navier-Stokes equations.
It consists in the flow of a two-dimensional (2D) fluid inside an infinite rectangular strip bounded by two walls.
Recently, this type of electronic flow has been experimentally realized and mapped spatially using different scanning-probe techniques in graphene samples \cite{sulpizio_visualizing_2019, ella_simultaneous_2019, ku_imaging_2020, huang_electronic_2023}. 

The plane Poiseuille solution respects the Navier-Stokes equations for any value of the average velocity of the flow.
For velocities below a certain critical threshold the plane Poiseuille flow is also stable, meaning that every small perturbation eventually decays, while, for velocities above this threshold, unstable perturbations\Add{,} that grow exponentially over time\Add{,} appear in the form of Tollmien-Schlichting traveling waves \cite{drazin_introduction_2002}. 
For even larger velocities the flow becomes turbulent.

The study of the stability of plane Poiseuille flow was initiated by Orr \cite{orr_stability_1907} and Sommerfeld \cite{sommerfeld_beitrag_1909} at the beginning of 20th century, but accurate solutions to the problem became available only with the use of computers \cite{thomas_stability_1953, orszag_accurate_1971}.
The plane Poiseuille flow is now known to become unstable when the Reynolds number ${\rm Re}^*=Wv^*/(2\nu)$, where $W$ is the width of the channel, $v^*$ is the velocity in the center of the stream and $\nu$ is the kinematic viscosity, is above a certain critical value ${\rm Re}^* = 5772$ \cite{orszag_accurate_1971}.

Electronic hydrodynamic instabilities have been studied starting from the work of Dyakonov and Schur \cite{dyakonov_shallow_1993}, that proposed them as a way to generate and detect TeraHertz radiation.
Different theoretical proposals have been presented \cite{mendoza_preturbulent_2011, furtmaier_rayleigh-benard_2015, gabbana_prospects_2018} but little conclusive experimental evidence has been shown insofar.

In this work we address the question of the stability of the Poiseuille electronic flow in a two-dimensional conductor, by considering the peculiarity that differentiate the electronic Poiseuille flow from the Poiseuille flow for normal fluids.
We also discuss the feasibility of an experiment to probe the existence of this instability. 

To this aim in Sect.~\ref{sect:equations} we summarize the relevant fluid mechanics equations for electronic systems. 
In Sect.~\ref{sect:stability} we develop a linear stability analysis for our problem leading to a modified Orr-Sommerfeld equation \cite{orr_stability_1907, sommerfeld_beitrag_1909} that is solved numerically.
The numerical results are presented in Section ~\ref{sect:results}.
Section~\ref{sec:conclusions} contains our conclusions and perspectives on future investigations.
The mathematical details of the calculations are presented in Appendices \ref{app:orr_sommerfeld}, \ref{app:eigenvalues}, 
\ref{app:poiseuille}, \ref{app:boundary}.
The numerical codes used in this study are available at \cite{noauthor_iacopo_2026}.
\section{Fluid dynamics equations for electronic flow}
\label{sect:equations}
In very clean, conducting, two-dimensional materials, the flow of electrons can be described, within a certain parameter window \cite{torre_nonlocal_2015, bandurin_negative_2016}, using the equations for an incompressible hydrodynamic flow \cite{torre_nonlocal_2015, levitov_electron_2016, bandurin_negative_2016}, namely the continuity equation and the Navier-Stokes equation.

The continuity equation for the \Add{incompressible electron flow} \Rem{electron velocity} is
\begin{equation}\label{eq:continuity}
     \nabla \cdot {\bm v}=0,
\end{equation}
where $\bm v$ is the velocity of the electronic fluid.
\Rem{The assumption of incompressible flow is well justified since the plasmon velocity, that plays the same role for electrons as the sound velocity for fluids, is much higher than the typical velocity of the flow in most practical situations} \cite{alonso-gonzalez_acoustic_2017, barcons_ruiz_experimental_2023}\Rem{.}

\Add{The assumption of incompressible flow is well justified in most experiments involving 2D materials since the acoustic plasmon velocity, that plays the same role for electrons as the sound velocity for fluids, is typically much higher than the velocity of the electronic flow.
In the hydrodynamic regime the acoustic plasmon velocity of a 2D electron fluid is always larger than $\vf/\sqrt{2}$ \cite{torre_acoustic_2019}, $\vf$ being the Fermi velocity, and, to the best of our knowledge, the lowest plasmon velocities reported experimentally are even larger than $\vf$ \cite{alonso-gonzalez_acoustic_2017, barcons_ruiz_experimental_2023}.
On the other side the average velocity of the flow is normally much smaller than $\vf$, allowing the ratio between the average velocity and the plasmon velocity to stay below $0.3$ as normally required for the flow to be considered incompressible.}

\Add{However, since the critical velocities for instability predicted by our calculations are very high, the assumption of incompressible flow should be reexamined {\it a posteriori} before applying our results to any concrete experimental situation.}

The electronic Navier-Stokes equation reads
\begin{equation}\label{eq:navier_stokes}
    D_{t}{\bm v}=\frac{e}{m}\nabla \Phi+ \nu\nabla ^{2}{\bm v}-\frac{1}{\tau}{\bm v} .
\end{equation}
Here,  $D_t=\partial_t + (\bm v\cdot \nabla)$ is the material derivative, $e$ is the electron charge, $m$ is the electron effective mass \cite{torre_nonlocal_2015}, $\Phi$ is the electrochemical potential, $\nu$ is the kinematic viscosity of the electron liquid \cite{principi_bulk_2016, krishna_kumar_superballistic_2017}, and $\tau$ is the momentum relaxation time \cite{torre_nonlocal_2015} that can be extracted from mobility measurements in wide samples where boundary effects are negligible.

This equation is identical to the Navier-Stokes equation that expresses the momentum balance  for an incompressible fluid (with the potential gradient replacing the pressure gradient), apart from the last term that represents the momentum of the electronic fluid that is lost due to collision with phonons and defects of the material.
This last term is formally equivalent to the Darcy \cite{darcy_fontaines_1856, avramenko_investigation_2005} term that is used in describing the flow of a fluid in a porous material.

It is worth noting that in Eq.~\eqref{eq:navier_stokes} we assumed that the electrons in our 2D material have a finite effective mass $m^*$. 
This includes the case of electrons in high mobility GaAs/AlGaAs quantum wells that have been shown to display hydrodynamic behavior in a certain range of parameters \cite{de_jong_hydrodynamic_1995, braem_scanning_2018}, but does not apply directly to single-layer graphene.
Indeed, the pseudo-relativistic Hamiltonian of single-layer graphene leads to a different form of the convective derivative that have been shown to have a quantitative impact on hydrodynamic instabilities of the Dyakonov-Schur type \cite{tomadin_theory_2013}.

Moreover, close to the charge neutrality point, both in single and bi-layer graphene, electrons and holes coexist, leading to the necessity of a two-fluid description \cite{narozhny_hydrodynamics_2015, narozhny_hydrodynamic_2017}.  
The study of  the stability of the plane-Poiseuille flow in the specific case of graphene, taking into account both the pseudo-relativistic convective derivative and the presence of electrons and holes, will be the subject of a future study.
We believe, however, that the conclusions of this study hold qualitatively also for graphene.

To solve the Navier-Stokes equation we need boundary conditions on the normal and tangential components of the velocity at the boundary.
The boundary condition of the normal component establishes that no current can leave the sample through the edges and reads
\begin{equation}\label{eq:bc_normal}
   \hat{\bm n}(\bm r)\cdot \bm{v}(\bm r) =0,
\end{equation}
where $\hat{\bm n}(\bm r)$ is the outward normal to the system boundary at the position $\bm r$.
This condition applies when $\bm r$ is on the boundary of the sample, excluding the regions where a lateral contact \cite{wang_one-dimensional_2013} is present.

The tangential boundary condition describes the friction produced by the sample boundaries on the electronic flow.
In electron hydrodynamics it is customary to assume a Robin boundary condition  in the form \cite{torre_nonlocal_2015, pellegrino_nonlocal_2017, moessner_boundary-condition_2019, kiselev_boundary_2019}
\begin{equation}\label{eq:bc_tangential}
\begin{split}
  \hat{n}_i(\bm r)\epsilon_{ij}  \left[\partial_jv_k(\bm r)+\partial_kv_j(\bm r)-\delta_{jk} \partial_\ell v_\ell(\bm r)\right]\hat{n}_k(\bm r)=\\
  =-\frac{1}{\lb}\epsilon_{ij}\hat{n}_i(\bm r)v_j(\bm r),
\end{split}
\end{equation}
where $\epsilon_{ij}$ is the 2D Levi-Civita tensor and the parameter $\lb>0$ is known as boundary scattering length \cite{torre_nonlocal_2015, pellegrino_nonlocal_2017}.
This boundary condition expresses a direct proportionality between the tangent velocity component and the force exerted by the boundary on the fluid in the tangential direction.
Mathematically, Eq.~\eqref{eq:bc_tangential} interpolates between the no-slip boundary condition (for $\lb=0$) and the free-surface boundary condition (for $\lb \to\infty$).
The parameter $\lb$ is related to the degree of specular scattering of electrons at the boundary of the sample \cite{pellegrino_nonlocal_2017} and depends on the roughness of the edges.
Recent experiments \cite{keser_geometric_2021} have explored the regime of very large $\lb$ using extremely clean edges.
\section{Stability of Poiseuille flow}
\label{sect:stability}
\begin{figure}[ht!!]
    \centering
    \begin{overpic}{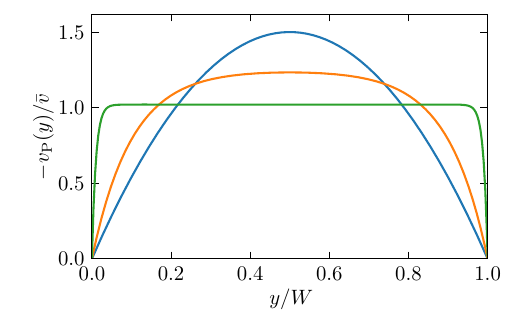}
        \put(0,55){$(a)$}
        \put(33,20){\frame{\includegraphics[scale=0.55]{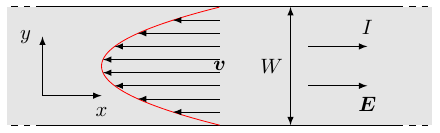}}}
    \end{overpic}
    \begin{overpic}{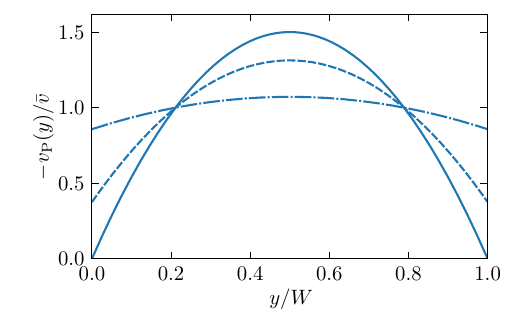}
        \put(0,55){$(b)$}
    \end{overpic}
    \caption{(Color online) (a) Flow velocity in the channel as a function of the transverse coordinate $y$, normalized to the average velocity $\bar{v}$, for different values of the ratio $W/D$. The blue line corresponds to $W/D\to0$ (standard plane-Poiseuille flow), the orange line to $W/D=10$, and the green line to $W/D=100$. In all cases $\lb=0$. The inset shows the sample geometry with the directions of electric field, electric current, and velocity of the electronic fluid. (b) Same as in (a) for $W/D\to0$ (standard plane-Poiseuille flow) and different values of the boundary scattering length $\lb$. The solid line corresponds to $\lb=0$, the dashed line to $\lb=0.1W$ and the dash-dotted line to $\lb =W$.}
    \label{fig:poiseuille}
\end{figure}
In this work we address the stability of the plane-Poiseuille electronic flow.
This is a stationary solution of the equations \eqref{eq:continuity}-\eqref{eq:navier_stokes} in an infinite two-dimensional strip of width $W$ ($-\infty <x<\infty$, $0\leq y\leq W$), under the action of a constant electric field in the direction of the strip, corresponding to a potential $\Phi_0 =-E_x x$, and with the boundary conditions \eqref{eq:bc_normal}-\eqref{eq:bc_tangential}.
A sketch of the sample geometry is shown in the inset of Fig.~\ref{fig:poiseuille}-a.

The velocity field corresponding to Poiseuille flow has the form $\bm v(\bm r, t)= \hat{\bm x}v_{\rm P}(y)$, where $v_{\rm P}(y)$ can be calculated by integrating the corresponding ordinary differential equation \cite{torre_nonlocal_2015}, yielding
\begin{equation}\label{eq:poiseuille_profile}
    v_{\rm P}(y) = -v_0 \left[1-\frac{\cosh\left(\frac{y-W/2}{D}\right)}{\cosh\left(\frac{W}{2D}\right)+\frac{\lb}{D} \sinh\left(\frac{W}{2D}\right)}\right].
\end{equation}
Here, $v_0=e\tau E_x/m=\mu E_x$ is the equilibrium velocity of the fluid in absence of boundary effects (i.e. in the bulk), $\mu$ being the electronic mobility $\mu = e\tau/m$, $D=\sqrt{\nu \tau}$ is the Gurzhi diffusion length \cite{torre_nonlocal_2015, levitov_electron_2016, bandurin_negative_2016}, $W$ is the width of the strip and $y$ the transverse coordinate.

Figure \ref{fig:poiseuille} shows examples of the Poiseuille velocity profile for different values of the parameters.
For $D\gg W$ the velocity profile reduces to the pure Poiseuille flow with a parabolic shape. 
On the other side, for $D\ll W$ the flow reduces to a uniform flow with two well-defined boundary layers of width $D$ at the boundaries.

Panel \ref{fig:poiseuille}-b shows the impact of the boundary scattering length $\lb$.
As $\lb$ is increased from zero a finite flow velocity is allowed near the edges making the flow profile more uniform.
In the limit $\lb\gg W$ the impact of boundary friction disappears and the flow becomes completely uniform across the section of the channel.
As we will demonstrate this makes the flow more stable and makes the realization of electronic flow instabilities more difficult. 

The velocity $v_0$ bears a simple relation with the applied electric field $E_x$ but is difficult to measure directly.
For this reason we prefer to use as characteristic velocity of the flow the absolute value of the average velocity 
\begin{equation}\label{eq:average_velocity}
\begin{split}
    \bar{v}=-\frac{\int_0^W  v_{\rm P}(y)dy}{W}
     =v_0 \frac{1+\left(\frac{\lb}{D}-\frac{2D}{W}\right)\tanh\left(\frac{W}{2D}\right)}{1+\frac{\lb}{D} \tanh\left(\frac{W}{2D}\right)}.
\end{split}
\end{equation}
that is directly related to the total electric current $I$ flowing in the sample by $I=e\bar{v}nW$, $n$ being the electronic density.

The Poiseuille flow described by \eqref{eq:poiseuille_profile} is a solution of the hydrodynamic equations \eqref{eq:continuity}, \eqref{eq:navier_stokes} that fulfills the correct boundary conditions \eqref{eq:bc_normal} \eqref{eq:bc_tangential} for any value of the average velocity $\bar{v}$ or, equivalently, of the total electric current.
However, this solution may or may not be stable depending on the value of $I$.
The central result of this work is a linear stability analysis of the Poiseuille solution of the electronic Navier-Stokes equations.
This is done by constructing a solution of the Navier-Stokes equations given by the sum of the Poiseuille solution plus a small perturbation $\delta {\bm v}$ in the form \footnote{The traveling wave ansatz of the perturbation introduced in \eqref{eq:perturbed_solution} implicitly assumes periodic boundary conditions in the $x$-streamwise direction of the flow.}
\begin{equation}\label{eq:perturbed_solution}
    \bm v(\bm r,t)= \hat{\bm x}v_{\rm P}(y) + \Re \left[\delta {\bm v}(y)e^{i(kx-\omega t)}\right],
\end{equation}
by substituting in the Navier-Stokes equations and retaining only the linear terms in the perturbation we can derive (See Appendix \ref{app:orr_sommerfeld} for details) an eigenvalue equation for the angular frequency $\omega$ of the perturbation in the form of a modified Orr-Sommerfeld \cite{orr_stability_1907, sommerfeld_beitrag_1909} equation
\begin{widetext}
 \begin{equation}\label{eq:orr_sommerfeld}
-i\nu \delta v_y''''(y) 
+ k\left[2i\nu k -v_{\rm P}(y)\right]\delta v_y''(y)
+k\left[v_{\rm P}''(y)+k^2v_{\rm P}(y) -i\nu k^3\right]\delta v_y(y)=
\left(\omega +\frac{i}{\tau}\right) \left[k^2 \delta v_y(y)-\delta v_y''(y)\right].
\end{equation}   
\end{widetext}
Enforcing the boundary condition \eqref{eq:bc_normal} on the perturbed velocity field leads to the boundary conditions on the $y$ component of the perturbation
\begin{equation}\label{eq:bc_normal_perturbation}
    \delta v_y(y=0,W)=0,
\end{equation}
while from \eqref{eq:bc_tangential} we obtain two more boundary conditions on $\delta v_y$
\begin{equation}\label{eq:bc_tangential_perturbation}
    \delta v_y'(y=0,W)=\pm\lb \delta v_y''(y=0,W),
\end{equation}
where the upper (lower) sign applies to $y=0$ ($y=W$).

Equation \eqref{eq:orr_sommerfeld} differs from the original Orr-Sommerfeld equation \cite{orr_stability_1907, sommerfeld_beitrag_1909} for plane-Poiseuille flow because of 
(i) the presence of the relaxation term $\tau^{-1}$, 
(ii) the modified Poiseuille profile $v_{\rm P}(y)$,
(iii) the presence of a finite $\lb$ in the boundary condition \eqref{eq:bc_tangential_perturbation}.
Moreover, the Orr-Sommerfeld equation is often expressed using a Lagrange stream function \cite{thomas_stability_1953} to represent the flow as $v_i =\epsilon_{ij} \partial_j \psi$. 
Our approach is equivalent as the perturbation of the stream function is proportional to the perturbation of $v_y$ as $\delta v_y(y) =-ik\delta\psi(y)$.

The original plane-Poiseuille problem has only one length scale, that is the width of the channel $W$, while in our problem there are two additional length scales $D$ and $\lb$.
Using the width $W$ of the channel as a unit of length and the average velocity $\bar{v}$ as unit of velocity, the problem depends on three dimensionless parameters: the ratios $D/W$ and $\lb/W$, and the Reynolds number 
\begin{equation}\label{eq:Reynolds_definition}
     \rey = \frac{\bar{v}W}{\nu} = 0.62\frac{I[{\rm mA}]}{\nu[{\rm m^2/s}] n[10^{12}~{\rm cm^{-2}} ]}.
\end{equation}
We choose this definition of the Reynolds number to preserve the proportionality between $\rey$ and the electric current that is the most relevant experimental quantity.
However, this differs from the Reynolds number definition ${\rm Re}^*$ that is commonly used in the fluid mechanics literature on plane Poiseuille flow\cite{thomas_stability_1953, orszag_accurate_1971}.
For the sake of comparison, the explicit form of the conversion factor is
\begin{equation}\label{eq:conversion_reynolds}
    {\rm Re}^*= \frac{1}{2}\frac{\cosh\left(\frac{W}{2D}\right)+\frac{\lb}{D}\sinh\left(\frac{W}{2D}\right) -1}{\cosh\left(\frac{W}{2D}\right)+\left(\frac{\lb}{D}-\frac{2D}{W}\right)\sinh\left(\frac{W}{2D}\right)}\rey.
\end{equation}
In the special case of pure plane-Poiseuille flow ($D\gg W$) with no-slip boundary condition ($\lb =0$) Eq.~\ref{eq:conversion_reynolds} simplifies to ${\rm Re}^*= 3\rey/4$.

To solve numerically \eqref{eq:orr_sommerfeld} we expand the solution on a set of basis functions that respect the correct boundary conditions \eqref{eq:bc_normal_perturbation}-\eqref{eq:bc_tangential_perturbation}. 
We use as basis function on the interval $0\leq \y\leq 1$, where $\y=y/W$, the functions
\begin{equation}\label{eq:basis_functions}
    F_n(\y) = \left[\y(1-\y) +2\frac{\lb}{W}\right]\sin(n\pi \y).
\end{equation}
This procedure, as detailed in Appendix~\ref{app:eigenvalues} leads to the generalized eigenvalue problem 
\begin{equation}\label{eq:eigenvalue_problem}
    L(\rey, kW, D/W, \lb/W)\cdot \bm u_n = \lambda_n  B(kW,\lb/W) \cdot \bm u_n ,
\end{equation}
where the explicit expressions of the matrices $L$ and $B$ are given in Appendix ~\ref{app:eigenvalues}.
The infinite matrices appearing in \eqref{eq:eigenvalue_problem} are truncated to a size $N\times N$ and the eigenvalues can be found numerically.
The problem is further simplified by the use of symmetry that makes all the matrix elements coupling functions of different parity vanish.
\Rem{In this work we focus only on symmetric perturbations (corresponding to $F_n$ with $n$ odd) as it is well known \cite{orszag_accurate_1971} that instability in plane-Poiseuille flow first appears in the symmetric sector. }
\Add{In this work we focus on symmetric perturbations (sinuous modes), corresponding to $F_n$ with $n$ odd, as it is well known \cite{orszag_accurate_1971, drazin_introduction_2002} that instability in plane-Poiseuille flow appears in the symmetric sector of the function space
\footnote{We analyzed also de antisymmetric sector and found no unstable modes for $\rey<10^7$ when $W/D<15$. 
An antisymmetric (varicose) unstable mode appears for $W/D>15$ but it occurs at higher values of $\rey$ than the sinuous mode and it is constituted essentially by an odd combination of edge modes that becomes degenerate with the corresponding even combination for $W/D\to \infty$.}.}

The angular eigenfrequencies of the perturbations are obtained from 
\begin{equation}\label{eq:eigenfrequencies}
    \omega_n = 2\pi \bar{f}\left(\lambda_n-i\frac{W^2}{\rey D^2}\right),
\end{equation}
where $\lambda_n$ are the eigenvalues of \eqref{eq:eigenvalue_problem} and $\bar{f}$ is a frequency scale given by
\begin{equation}\label{eq:frecuency_scale}
    \bar{f}=\frac{\bar{v}}{2\pi W} = 0.099 ~{\rm THz } \frac{I[{\rm mA}]}{(W[{\rm \mu m}])^2 n[10^{12}~{\rm cm^{-2}} ]}.
\end{equation}
The flow is unstable if one perturbation with $\Im [\omega]>0$  exists, so it is sufficient to sort the obtained eigenvalues by decreasing imaginary part and check the sign of the imaginary part of the most unstable eigenfrequency $\omega_0$.
\section{Results}
\label{sect:results}

\begin{figure}[ht!!]
    \centering
    \begin{overpic}{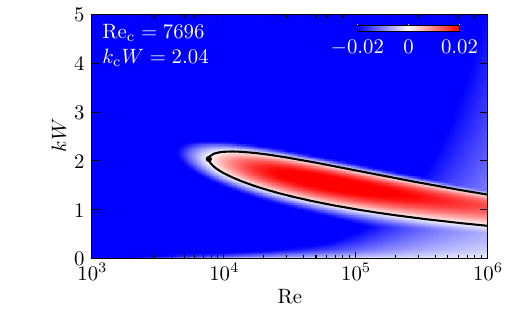}
        \put(0,55){$(a)$}
    \end{overpic}
    \begin{overpic}{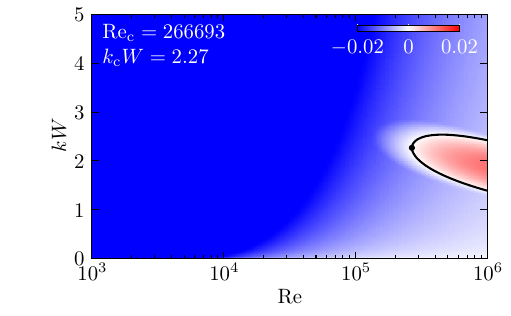}
        \put(0,55){$(b)$}
    \end{overpic}
    \begin{overpic}{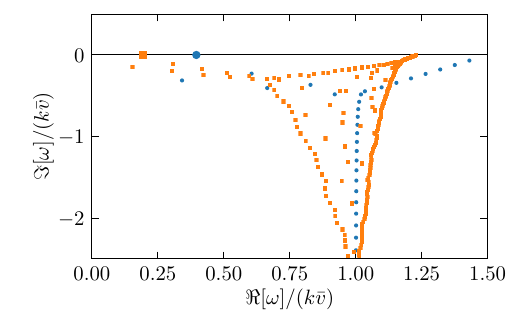}
        \put(0,55){$(c)$}
    \end{overpic}
    \caption{(Color online) (a) Stability diagram of electronic Poiseuille flow. The color plot represents the imaginary part of the most unstable frequency as a function of the Reynolds number and of the wavenumber $k$ of the perturbation. The black line separates the region where $\Im [\omega]<0$ (stable) from the region where $\Im [\omega]>0$ (unstable). The black dot marks the position of the smallest Reynolds number where instability occurs.
    (b) Same as (a) for $W/D=10$ and $\lb=0$.
    (c) Complex-plane representation of the eigenvalues of the Orr-Sommerfeld equation at the critical point ($\rey=\reyc$, $k=\kc$) for $W/D\to 0$ (blue dots) and $W/D=10$ (orange squares). The larger symbols highlight the eigenvalues that are crossing the real axis.}
    \label{fig:stability_k}
\end{figure}

To assess the stability of Poiseuille flow we solve numerically the eigenvalue problem \eqref{eq:eigenvalue_problem} fixing the values of $D/W$ and $\lb/W$ and varying the Reynolds number $\rey$ and the wavevector $k$ of the perturbation.
An example of the results is shown in Figure~\ref{fig:stability_k}-a,b where the stability diagram in the $\rey - kW$ plane is shown for the limiting case \Rem{$D\gg W$}\Add{$W\ll D$}, that corresponds to the standard plane Poiseuille flow (See Appendix ~\ref{app:poiseuille} for details of this limiting case), and for \Rem{$D/W=0.1$}\Add{$W/D=10$}.
The color plot represents the imaginary part of the eigenfrequency $\omega_0$ associated to the most unstable perturbation with blue denoting negative imaginary part (stable) and red denoting positive imaginary part (unstable).
The black contour curve is the neutral stability boundary that separates the regions of stable and unstable perturbations.
For low Reynolds numbers, small perturbations eventually decay but, if the Reynolds number (the current) increases above a critical value, $\reyc$, perturbations with streamwise periodicity $2\pi/\kc$, $\kc$ marked as a black dot in the diagram, exponentially grow. 
For $\rey>\reyc$ a continuous spectrum of wave-like perturbations may also grow and destabilize the base flow.
By comparing the two diagrams we see that decreasing $D$ leads to a more stable flow, resulting in a higher critical Reynolds number and a higher current needed to destabilize the base flow.
Figure ~\ref{fig:stability_k}-c shows the spectra of eigenfrequencies in the complex plane at the critical point ($\rey=\reyc$, $k=\kc$) for the same values of $D/W$.
\begin{figure}[ht!!]
    \centering
    \begin{overpic}{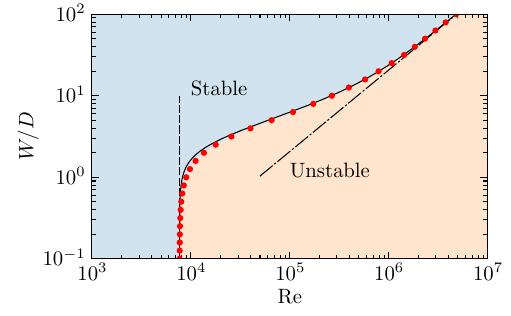}
        \put(0,55){$(a)$}
    \end{overpic}
    \begin{overpic}{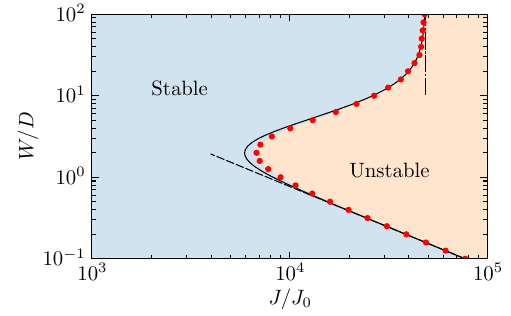}
        \put(0,55){$(b)$}
    \end{overpic} 
    \begin{overpic}{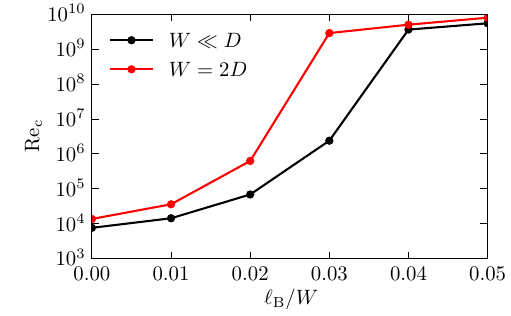}
        \put(0,55){$(c)$}
    \end{overpic} 
    \caption{(Color online) (a) Stability diagram as a function of $\rey$ and $W/D$. Red dots represent the numerical values of $\reyc$. 
    The black dashed line represents the Poiseuille limit ($W\ll D$) $\reyc = \reycz$. 
    The black dash-dotted line represents the $W\gg D$ limit $\reyc=\reyci W/D$. The black thin line represents the approximate expression \eqref{eq:reynolds_approx}.  
    (b) Same as (a) expressed as a function of the current density in units of $J_0$ on the horizontal axis.
    (c) Dependence of the critical Reynolds number on $\lb$ for $W/D\ll 1$ (black symbols) and $W/D=2$, corresponding to the minimum of $J/J_0$ in (b), (red symbols).
    The dots represent the numerical points, the solid line is a guide to the eye.}
    \label{fig:phase_diagram}
\end{figure}

The most fundamental question we would like to answer with this work is whether it is possible to observe electronic instabilities in a plane Poiseuille geometry and to quantify the value of the current at which the onset of these instabilities can be expected.
To this aim we apply the same procedure to different values of $D/W$ and $\lb/W$.
We start from $\lb=0$ since we expect this case to be the most favorable for the turbulence.
The numerical data for the critical Reynolds number are given in Table~\ref{tab:results}.

Figure ~\ref{fig:phase_diagram}-a shows the region of stability in terms of the Reynolds number and the ratio $W/D$.
At low values of $W/D$ the critical $\rey$ reduces to the value of plane-Poiseuille flow $\reycz=\reycznum$.
Increasing the value of $W/D$ always increases $\reyc$, and for very high values of $W/D$ the critical Reynolds number starts to behave linearly with $W/D$ according to the approximate law $\reyc=\reyci W/D$ with $\reyci\approx \reycinum$.
This linear behavior is due to the decoupling of the two edges when $W\gg D$.
When $W\gg D$ the two edges of the conductor act independently and the velocity profile close to the edge at $y=0$ is 
\begin{equation}\label{eq:boundary_layer_flow}
    v_{\rm P}(y)\approx -v_0[1-e^{-y/D}/(1+\lb/D)].
\end{equation}
The stability of this flow profile is controlled by a Reynolds number $\reyi = Dv_0/\nu$ defined in terms of the distance $D$, as $W$ effectively drops out of the problem, with instability arising when $\reyi$ is higher than a critical values $\reyci$.
Using $\reyi = (Dv_0)/(W\bar{v})\rey$ and $\bar{v}\to v_0$ for $W\gg D$ we obtain an asymptotic linear behavior of the critical Reynolds number with angular coefficient $\reyci$.
The modified Orr-Sommerfeld equation \eqref{eq:orr_sommerfeld} with the velocity profile \eqref{eq:boundary_layer_flow} for $y>0$ is solved in Appendix \ref{app:boundary} and yields $\reyci\approx \reycinum$. 
By the same reasoning it is possible to show that for $W\gg D$ also the values of $\kc W$ and of the frequency of the first unstable perturbation \Add{$\fc/\bar{f}$} \Rem{$f_{\rm cr}/\bar{f}$} depend linearly on the ratio $W/D$.
The proportionality coefficients are shown in the last line of Table~\ref{tab:results}.

To interpolate between these two limiting behaviors we introduce a simplified expression for the critical Reynolds number 
\begin{equation}\label{eq:reynolds_approx}
\reyc\approx \reycz + \reyci \frac{\left(\frac{W}{D}\right)^3}{\left(\frac{W}{D}\right)^2+a^2}.    
\end{equation}
A best fit to the data gives $a\approx \ainterp$.
This simplified expression can be used to calculate approximately the critical Reynolds number without performing any numerical calculation.

Our definition of the Reynolds number is proportional to the electric current flowing in the channel.
However, in order to decide the feasibility of an experiment probing the instability of electronic Poiseuille flow it is more interesting to express the results in terms of the 2D current density flowing in the channel as the latter is the real limiting factor in experiments.
The current density is given by $J= I/W$ and can be expressed as $J=\rey D/W J_0$, where
\begin{equation}\label{eq:current density}
    J_0 =\frac{en\nu}{D}=1.6 ~\frac{{\rm mA}}{{\rm \mu m}} n[10^{12}~{\rm cm^{-2}} ]\sqrt{\frac{\nu[{\rm m^2/s}]}{\tau[{\rm ps}]}}.
\end{equation}
Figure ~\ref{fig:phase_diagram}-b shows the stability diagram in terms of the current density $J/J_0$.
From the figure we can see that the current density needed to induced turbulence has indeed a minimum at $W\approx \Wmin D$ with a minimum value $J_{\rm min}\approx \jmin ~ J_0$.
This represents the most favorable parameter setting for an experiment aimed at probing electronic Poiseuille instability.

Finally, Figure ~\ref{fig:phase_diagram}-c highlights the impact of a finite boundary scattering length $\lb$ for the case $W\ll D$ and $W=2D$.
As expected, for a given value of $D/W$, increasing $\lb$ makes the flow more uniform and more stable, and raises the $\reyc$ by several order of magnitude even with relatively small values of $\lb/W$.

The linear stability analysis allows to characterize the spatio-temporal structure of the instability.
In particular it allows to calculate the wavenumber $\kc$, the frequency and the spatial shape of the first unstable perturbation, i.e., of the eigenmode that crosses $\Im [\omega] = 0$ at the critical point.
The frequency of the perturbation is obtained from the real part of the most unstable eigenvalue at the critical point.
The spatial form of the perturbation is obtained from the eigenvector associated to the same eigenvalue and the basis functions \eqref{eq:basis_functions}.

\begin{figure}[ht!!]
    \centering
    \begin{overpic}{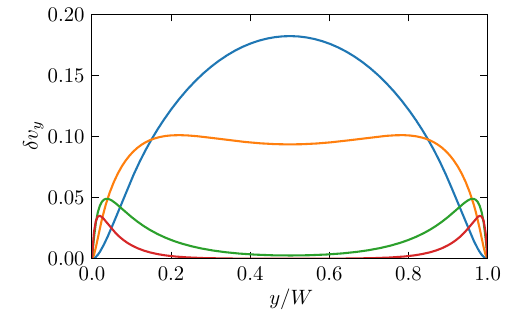}
        \put(0,55){$(a)$}
    \end{overpic}
    \begin{overpic}{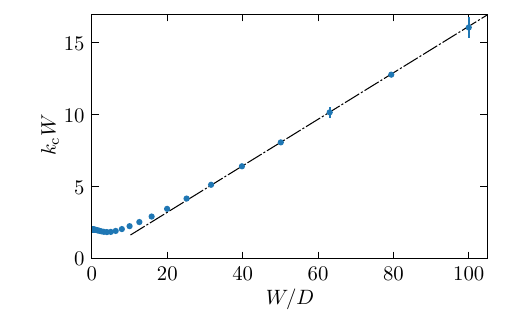}
        \put(0,55){$(b)$}
    \end{overpic}
    \begin{overpic}{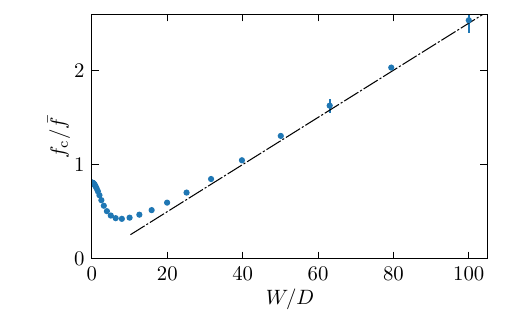}
        \put(0,55){$(c)$}
    \end{overpic}
    \caption{(Color online) (a) Real part of the Velocity profile of the most unstable perturbation at the critical point for different values of the ratio $W/D$: $W/D \ll 1$ (blue line), $W/D =10$ (orange line), \Rem{$W/D =40$} \Add{$W/D =50.12$} (green line), and $W/D =100$ (red line). 
    The phase of the eigenfunctions is chosen to have real first component. The imaginary (out of phase) parts are too small to be represented in this figure.
    (b) Critical wavenumber as a function of $W/D$ for $\lb=0$. The dots with error bars represent numerical value, the dash-dotted line corresponds to the $W\gg D$ limit.  
    (c) Frequency of the critical perturbation in units of $\bar{f}$ as a function of $W/D$ for $\lb=0$. The dots with error bars represent numerical value, the dash-dotted line corresponds to the $W\gg D$ limit.   }
    \label{fig:critical}
\end{figure}

Figure ~\ref{fig:critical}-a shows the spatial shape of the first unstable perturbation for different values of $W/D$ and $\lb=0$.
The transition from a center mode for $W\ll D$ to a wall mode for $W\gg D$ can be observed.

Figures ~\ref{fig:critical}-b-c show the frequency (in units of the frequency scale \ref{eq:frecuency_scale} ) of the first unstable perturbation and the corresponding wavevector.
In the most favorable situation, i.e., when $W\approx 2D$, the instability is expected to start in the form of an exponentially growing perturbation with wavenumber $\kc=\kcmin/W$ and frequency $f=\fmin \bar{f}$.
\begin{table*}
\begin{tabular}{ccccc}
\hline
\hline
$W/D$ & $N$ & $\reyc$ & $\kc W$ & $\fc/\bar{f}$\\
\hline
$0$ & $1000$ & $7.6961(2) \times 10^{3}$ & $2.0402(9) $ & $0.8078(5) $\\
$0.1$ & $1000$ & $7.7084(2) \times 10^{3}$ & $2.04077(2) $ & $0.80786(2) $\\
$0.1259$ & $1000$ & $7.71569(5) \times 10^{3}$ & $2.04060(4) $ & $0.80761(3) $\\
$0.1585$ & $1000$ & $7.72707(4) \times 10^{3}$ & $2.0404(2) $ & $0.80727(6) $\\
$0.1995$ & $1000$ & $7.74508(8) \times 10^{3}$ & $2.03978(2) $ & $0.80657(2) $\\
$0.2512$ & $1000$ & $7.7736(2) \times 10^{3}$ & $2.0396(6) $ & $0.8059(3) $\\
$0.3162$ & $1000$ & $7.8192(2) \times 10^{3}$ & $2.0376(3) $ & $0.8039(2) $\\
$0.3981$ & $1000$ & $7.8917(3) \times 10^{3}$ & $2.0359(3) $ & $0.8016(2) $\\
$0.5012$ & $1000$ & $8.00772(4) \times 10^{3}$ & $2.0333(2) $ & $0.79794(8) $\\
$0.631$ & $1000$ & $8.19364(4) \times 10^{3}$ & $2.02884(7) $ & $0.79198(4) $\\
$0.7943$ & $1000$ & $8.49394(7) \times 10^{3}$ & $2.02181(7) $ & $0.78272(4) $\\
$1$ & $1000$ & $8.98408(8) \times 10^{3}$ & $2.0113(1) $ & $0.76869(5) $\\
$1.259$ & $1000$ & $9.79627(6) \times 10^{3}$ & $1.99559(5) $ & $0.74768(3) $\\
$1.585$ & $1000$ & $1.117068(8) \times 10^{4}$ & $1.9736(3) $ & $0.7175(2) $\\
$1.995$ & $1000$ & $1.35602(3) \times 10^{4}$ & $1.9438(3) $ & $0.6760(2) $\\
$2.512$ & $1000$ & $1.78398(3) \times 10^{4}$ & $1.908180(3) $ & $0.6232643(5) $\\
$3.162$ & $1000$ & $2.56788(5) \times 10^{4}$ & $1.8743(2) $ & $0.56387(5) $\\
$3.981$ & $1000$ & $4.0050(2) \times 10^{4}$ & $1.85528(8) $ & $0.50608(3) $\\
$5.012$ & $1000$ & $6.5531(4) \times 10^{4}$ & $1.86983(1) $ & $0.459824(8) $\\
$6.31$ & $1000$ & $1.0784(1) \times 10^{5}$ & $1.93556(5) $ & $0.431962(7) $\\
$7.943$ & $1000$ & $1.7286(4) \times 10^{5}$ & $2.06544(4) $ & $0.42474(3) $\\
$10$ & $1000$ & $2.666929(8) \times 10^{5}$ & $2.2680(2) $ & $0.43728(3) $\\
$12.59$ & $1000$ & $3.955(3) \times 10^{5}$ & $2.5544(4) $ & $0.4690(2) $\\
$15.85$ & $1000$ & $5.7765(2) \times 10^{5}$ & $2.9367(5) $ & $0.51745(9) $\\
$19.95$ & $1000$ & $7.90(3) \times 10^{5}$ & $3.471(3) $ & $0.5969(9) $\\
$25.12$ & $1000$ & $1.072(3) \times 10^{6}$ & $4.186(6) $ & $0.704(2) $\\
$31.62$ & $1000$ & $1.43(3) \times 10^{6}$ & $5.14(3) $ & $0.848(8) $\\
$39.81$ & $1000$ & $1.84(3) \times 10^{6}$ & $6.43(4) $ & $1.047(9) $\\
$50.12$ & $1600$ & $2.33(3) \times 10^{6}$ & $8.09(3) $ & $1.306(8) $\\
$63.1$ & $1600$ & $2.96(8) \times 10^{6}$ & $10.2(5) $ & $1.63(8) $\\
$79.43$ & $1600$ & $3.8(2) \times 10^{6}$ & $12.8(2) $ & $2.03(3) $\\
$100$ & $1600$ & $4.8(3) \times 10^{6}$ & $16.1(7) $ & $2.5(2) $\\
$\infty$ & $500$ & $[4.8(2) \times 10^{4}]W/D$ & $[0.162(5)]W/D$ & $[0.025(1)]W/D$\\
\hline
\hline
\end{tabular}
\caption{Numerical values of the critical Reynolds number $\reyc$, of the critical wavevector $\kc$, and of the critical frequency $f_{\rm cr}$ as a function of the ratio $W/D$.
$N$ denotes the number of basis functions used in the calculations. 
\Add{The numbers inside the parenthesis represent the uncertainty on the last significant digit.}
Uncertainities are estimated from the differences between the results obtained with $N$ and $N/2$ basis functions.}
\label{tab:results}
\end{table*}
\section{Conclusions}
\label{sec:conclusions}
Our work \Rem{completely} characterizes the linear stability of electronic plane-Poiseuille flow in two dimensions \Add{under the assumptions that Eqs.~\eqref{eq:continuity} (incompressible flow), and \eqref{eq:navier_stokes} (Navier-Stokes) hold.} \Rem{and} \Add{It also} provides clear guidance for the design of experiments aimed at studying the onset of instabilities in this geometry.
The best situation to observe electronic turbulence with minimal current density is a geometry with $W/D = \Wmin$ and $\lb\approx 0$, where the onset of instability is predicted to occur at $J=\jmin ~J_0$.

Even in this most favorable case, the current density needed to observe instabilities in this geometry is indeed quite high, at least for the typical values of parameters normally encountered in 2D materials, exceeding the highest values of current density reported in the literature \cite{berdyugin_out--equilibrium_2022} $\approx 1{\rm mA}/{\rm \mu m}$.

\Add{Moreover, at the high values of current needed to generate the instability the average velocity of the fluid can be larger than the Fermi velocity $\vf$. 
The ratio $\bar{v}/\vf$ is given by
\begin{equation}\label{eq:ratiov_vf}
 \frac{\bar{v}}{\vf}  =  \frac{\rey D^2}{W^2}\frac{W}{\tau \vf}\approx 3400\frac{W}{\ell_p},
\end{equation}
where in the last equality we used the values for $W=2D$ and recognized the momentum relaxation length $\ell_p=\vf \tau$.
This means that only for extremely clean systems the ratio $\bar{v}/\vf$ can be kept smaller than one \footnote{If the Fermi liquid theory can be applied to the system the kinematic viscosity can be expressed as $\nu = 1/4 \vf^2 \tau_{\rm ee}$, with $\tau_{\rm ee}$ being the electron-electron scattering time. 
In this case the ratio $\bar{v}/\vf$ at the onset of the instability and for $W=2D$ is $\bar{v}/\vf=3400 \sqrt{(\tau_{\rm ee}/\tau)}$. 
This emphasize the need for a large timescale separation between $\tau$ and $\tau_{\rm ee}$ that means that system is deeply in the hydrodynamic regime.}.
While a drift velocity $\bar{v}$ larger than $\vf$ is in principle possible in a parabolic-band conductor the assumptions behind the hydrodynamic model may become questionable in this case and the model could not be completely consistent when $\bar{v}/\vf$ is larger than one . 
}

However, it must be noted that linear stability analysis is known to overestimate the Reynolds number needed to trigger instabilities.
It has long been established that two-dimensional Poiseuille channel flows admit finite-amplitude nonlinear traveling waves at significantly lower Reynolds numbers. 
Rather than arising from a linear instability of the base flow, these waves originate at fold (saddle-node) bifurcations, distorting the flow dynamics within the phase space. 
Such waves were initially identified under constant pressure-gradient configurations by \cite{ehrenstein_three-dimensional_1991}. 
More recently, they have been computed and tracked in constant mass-flux configurations by \cite{mellibovsky_mechanism_2015} and \cite{ayats_symmetry-breaking_2020}, who concluded that these exact
coherent hydrodynamic structures may undergo instabilities leading to streamwise localization. 
Analogous solutions may also exist in the present problem; however, confirming this would necessitate a fully nonlinear analysis. 
Such an investigation is beyond the scope of the current study and remains a subject for future research.

\Add{
It should also be noted that the present linear stability framework is purely temporal, assuming harmonic perturbations proportional to $\exp[i(kx-\omega t))]$, in an infinitely long or periodic domain. 
In a finite-length channel, the practical observability of these instabilities depends on whether they are absolute or convective: a purely convective instability may be swept out of the domain before growing appreciably. 
While formal classification of the primary linear mode could be achieved via the Briggs-Bers saddle-point criterion, applying linear spatio-temporal stability theory to Poiseuille-like open shear flows provides limited physical insight into the actual transition process. 
The true absolute or convective nature of these subcritical structures - dictated by the relative propagation speeds of their leading and trailing fronts - ultimately governs whether they can be observed in a finite domain. 
A quantitative characterization of these front dynamics requires direct numerical simulations of the full non-linear equations, which falls outside the scope of the present linear temporal analysis.
}

Equation \eqref{eq:current density} serves as a guide for selecting the best 2D system and regime of parameters to probe the transition to turbulence of Poiseuille flow.
In particular, low \Add{electronic} density, low viscosity, and high mobility all favor the detection of turbulence by lowering $J_0$.
On the other end, large values of $\lb$ have an extremely detrimental effect on the onset of instabilities because they make the basic flow more uniform and therefore more stable.
This means that samples with relatively rough edges should be preferred for this experiment.

\Rem{According to our results,} \Add{Due to its low electronic density, low viscosity \cite{muller_graphene_2009}, and high mobility} graphene close to the charge neutrality point, either single-layer or bi-layer, seem to be the ideal choice but our analysis does not apply  directly as hydrodynamic equations for these materials are more complicated due to the coexistence of electrons and holes that requires a two-fluid description and a relativistic convective derivative.
\Add{Recently, another mechanisms of flow instability due to the density-dependence of current dissipation has been theoretically predicted \cite{liong_spontaneous_2025} in graphene close to the neutrality point.}

It must be \Rem{noticed}\Add{observed} that the geometry we studied, while simple, does not favor instabilities because it relies only on boundary scattering to create flow inhomogeneities, hence the strong dependence on $\lb$.
The generalization of this study to include the peculiarities of graphene and the impact of different experimental geometries like the ones studied in \cite{gabbana_prospects_2018} will be the subject of a separate publication.

We remark that other Authors have solved the same mathematical problem, at least for $\lb=0$, in other context.
For example Lock \cite{lock_stability_1955} and later Takashima \cite{takashima_stability_1996} solved a very closely related problem for a charged fluid in presence of a magnetic field.
Avramenko et al. \cite{avramenko_investigation_2005} solved the same problem in the context of flow in a porous material where they also consider a Forchheimer term.
In the cases where the numerical results can be compared directly they all agree with our study.

We hope that our study can stimulate more theoretical and experimental research on electron hydrodynamic phenomena in 2D materials as these are one of the few real systems where a lot of prediction of 2D hydrodynamics can be actually experimentally tested.

\bibliographystyle{apsrev4-2}
\bibliography{references}
\begin{acknowledgments}
We acknowledge useful discussions with Francesca Ribas Prats and Riccardo Bertini.

I.T. acknowledges support from the Spanish Ministerio de Ciencia e Innovación (MCIN/AEI/10.13039/501100011033, grant PID2023-147469NB-C21), and from the Generalitat de Catalunya (grant 2021 SGR 01411). 

A.M. acknowledges support from the Ministerio de Ciencia, Innovación y Universidades, Agencia Estatal de Investigación, project PID2023-150029NB-I00 (MCIU/AEI/10.13039/501100011033/FEDER, UE), and from the Generalitat de Catalunya (grant 2021 SGR 00586). 
\end{acknowledgments}
\appendix
\section{Derivation of the modified Orr-Sommerfeld equation \texorpdfstring{\eqref{eq:orr_sommerfeld}}{}}
\label{app:orr_sommerfeld}
 To study the stability of electronic Poiseuille flow we perform a linear stability analysis \cite{landau_fluid_1987} by adding a perturbation $\delta {\bm v}$ to the Poiseuille velocity field $\bm v_0=\hat{\bm x}v_{\rm P}(y)$ and a perturbation $\delta \Phi$ to the electric potential $\Phi_0=-E_xx$. 
 We substitute the perturbed fields into \eqref{eq:continuity}-\eqref{eq:navier_stokes} and drop all the non-linear terms in the perturbation obtaining two equations for the time evolution of the perturbation
\begin{equation}\label{eq:continuity_perturbation}
     \nabla \cdot \delta \bm{v}=0 ,
\end{equation}
and
\begin{equation}\label{eq:navier_stokes_perturbation}
\begin{split}
     \partial_{t} \delta\bm{v}+\left( \bm{v}_0\cdot\nabla  \right)\delta\bm{v}+\left( \delta\bm{v}\cdot\nabla  \right)\bm{v}_0=\\
     =\frac{e}{m}\nabla \delta \Phi+\nu\nabla ^{2}\delta\bm{v}-\frac{1}{\tau}\delta\bm{v}.
\end{split}
\end{equation}
On the boundaries, the perturbation must respect the same homogeneous boundary conditions \eqref{eq:bc_normal}-\eqref{eq:bc_tangential} as the velocity. 
Because of the spatial (in the streamwise $x$ direction) and temporal translational invariance of the system we can choose a perturbation in the form  
\begin{align}
     \delta\bm{v}(\bm r,t)&=\exp(ikx-i\omega t)\delta\bm{v}(y),\label{eq:velocity_perturbation}\\
     \delta\Phi(\bm r,t)&=\exp(ikx-i\omega t)\delta\Phi(y).\label{eq:potential_perturbation}
\end{align}
Substituting \eqref{eq:velocity_perturbation} and \eqref{eq:potential_perturbation} into \eqref{eq:continuity_perturbation} and \eqref{eq:navier_stokes_perturbation} we obtain the system of equations
\begin{widetext}
\begin{align}
     \delta v'_{y}+ik\delta v_{x} & =0,\label{eq:system_1}\\
     \left[-i\omega+ik v_{\rm P}(y) \right]\delta v_{x}(y)+v'_{\rm P}(y)\delta v_{y}(y)& =\frac{eik}{m}\delta\Phi(y) -\left(\nu k^{2}+\frac{1}{\tau} \right)\delta v_{x}(y)+\nu \delta v''_{x}(y),\label{eq:system_2}\\
     \left[ -i\omega +ikv_{\rm P}(y)\right]\delta v_{y}(y)&=\frac{e}{m}\delta\Phi'(y) -\left( \nu k^{2}+\frac{1}{\tau}\right)\delta v_{y}(y)+\nu \delta v''_{y}(y).\label{eq:system_3}
\end{align}
\end{widetext}
We can transform this system into a single equation by deriving \eqref{eq:system_2}, substituting into \eqref{eq:system_3}, and making use of \eqref{eq:system_1}. 
This yields Equation \eqref{eq:orr_sommerfeld}.
\section{Details of the eigenvalue problem \texorpdfstring{\eqref{eq:eigenvalue_problem}}{}}
\label{app:eigenvalues}
As it is customary in fluid mechanics and computational physics we prefer to work with dimensionless equations in order to minimize the number of independent parameters. 
To this end we use the width of the strip $W$ as unit of length, and the average velocity of the Poiseuille flow $\bar{v}$ as unit of velocity. 
Using these units we can define the dimensionless coordinate $\tilde{y} =y/W$, the dimensionless wavenumber $\tilde{k}=kW$, and the dimensionless angular frequency $\tilde{\omega} = \omega W/\bar{v}$. 
We also define $\lbr = \lb/W$ and $\Lambda = W/D=\tilde{D}^{-1}$.
In dimensionless units \eqref{eq:orr_sommerfeld} becomes
\begin{equation}\label{eq:vy_dimensionless}
\begin{split}
a_4 \delta v_y''''(\tilde{y}) 
+\left[ a_{2}+b_{2}\frac{\cosh(\Lambda\tilde{y}-\Lambda/2)}{\cosh(\Lambda/2)}\right]\delta v_y''(\tilde{y})\\
+\left[a_{0}+b_{0}\frac{\cosh(\Lambda\tilde{y}-\Lambda/2)}{\cosh(\Lambda/2)}\right]\delta v_y(\tilde{y})\\
=\left(\tilde{\omega} +\frac{i}{\tilde{\tau}}\right) \left[\tilde{k}^2 \delta v_y(\tilde{y})-\delta v_y''(\tilde{y})\right],
\end{split}
\end{equation}
%
with coefficients given by
\begin{align}\label{eq:vy_coefficients_2}
a_4  & = -\frac{i}{\rey}, \\
a_{2}  & = \tilde{k}\left(c^{-1}+\frac{2i\tilde{k}}{\rey} \right),\\
b_{2} & = - \frac{\tilde{k}c^{-1}}{1  +\lbr\Lambda \tanh(\Lambda/2)},\\
a_{0} & = -\tilde{k}^3\left(c^{-1}+i\frac{\tilde{k}}{\rey} 
    \right
    ),\\
b_{0} & = \frac{c^{-1}\tilde{k}\left(\tilde{k}^2+\Lambda^2\right)}{1 +\lbr \Lambda \tanh(\Lambda/2)},
\end{align}
where
\begin{equation}\label{eq:c_definition}
    c = 1-\frac{ 2\tanh(\Lambda/2)}{\Lambda[1+\lbr\Lambda\tanh(\Lambda/2)]}.
\end{equation}
We express the velocity perturbation as a linear combination of the functions \eqref{eq:basis_functions} in the form $\delta v_y(\tilde{y}) =\sum_{n} c_nF_n(\tilde{y})$. 
Substituting into \eqref{eq:vy_dimensionless}, multiplying by $F_m(\tilde{y})$ and integrating on $\tilde{y}$ between $0$ and $1$ we obtain the generalized eigenvalue problem \eqref{eq:eigenvalue_problem}, where
\begin{align}
    L &= a_4M^{(4)}+a_{2}  M^{(2)}+ +a_{0}M^{(0)}    +b_{2} N^{(2)} +b_{0} N^{(0)},\label{eq:L_definition}\\
    B &= \tilde{k}^2M^{(0)}-M^{(2)},\label{eq:B_definition}
\end{align}
and the matrices $M^{(i)}$, and $N^{(i)}$ are given by 
\begin{align}\label{eq:matrix_elements}
    M^{(0)}_{mn}&= \int_0^1 F_{m}(\y)F_{n}(\y)d\y,\\  
    M^{(2)}_{mn}&=\int_0^1 F_{m}(\y)F_{n}''(\y)d\y,\\
    M^{(4)}_{mn}&= \int_0^1 F_{m}(\y)F_{n}''''(\y)d\y,\\
    N^{(0)}_{mn}&= \int_0^1 F_{m}(\y)\frac{\cosh[\Lambda(\y-1/2)]}{\cosh[\Lambda/2]}F_{n}(\y)d\y,\\
    N^{(2)}_{mn}&= \int_0^1 F_{m}(\y)\frac{\cosh[\Lambda(\y-1/2)]}{\cosh[\Lambda/2]}F_{n}''(\y)d\y,  
\end{align}
All the integrals involved in the above matrix elements can be solved analytically.
However, their explicit expressions are cumbersome and not reported here.
They can be found in \cite{noauthor_iacopo_2026} together with Python functions implementing them.

Identifying the eigenvalues $\lambda_n$ of \eqref{eq:eigenvalue_problem} with $\tilde{\omega_n} +i\tilde{\tau}^{-1}$ yields Eq.~\eqref{eq:eigenfrequencies}.
\section{Pure plane Poiseuille limit \texorpdfstring{($W/D\ll 1$)}{}}
\label{app:poiseuille}
The limit $\Lambda =W/D \to 0$ of the matrix $L$ in \eqref{eq:L_definition} is singular and must be treated with care.
Making use of the asymptotic expressions 
\begin{equation}
    c^{-1} = 
    \frac{12}{1+6\lbr} \Lambda^{-2}
    +\frac{6(30 \lbr^2+10\lbr+1)}{5(6\lbr+1)^2} +O(\Lambda^2),
\end{equation}
and
\begin{equation}
    N^{(i)} = M^{(i)}+ \Lambda^2 C^{(i)}+O(\Lambda ^4),
\end{equation}
where
\begin{align}
    C_{mn}^{(0)} = \frac{1}{2}\int_0^1 F_m(y)y(y-1)F_n(y) dy,\\
    C_{mn}^{(2)} = \frac{1}{2}\int_0^1 F_m(y)y(y-1)F_n''(y) dy,
\end{align}
we obtain the limiting expression of $L$ for $W\ll D$
\begin{equation}
\begin{split}
    L= \frac{-i}{\rey}M^{(4)}
    +\left(\frac{2ik^2}{\rey}+\frac{6\lbr k}{1+6\lbr}   \right)M^{(2)}\\
    + \left(\frac{-ik^4}{\rey}+12k\frac{1-\lbr k^2/2}{1+6\lbr}\right)M^{(0)}  \\ 
    -\frac{12k}{1+6\lbr} C^{(2)} +\frac{12k^3}{1+6\lbr} C^{(0)} +O(\Lambda^2).
\end{split}
\end{equation}
\section{Boundary layer limit \texorpdfstring{($W/D\gg 1$)}{}}
\label{app:boundary}
In this limit we need to solve Eq.~\eqref{eq:orr_sommerfeld} on the half-line $y>0$, with a base flow given by \eqref{eq:boundary_layer_flow}, and subjected to the boundary conditions
\begin{equation}\label{eq:bc_normal_boundarylayer}
    \delta v_y(y=0)=0,
\end{equation}
and
\begin{equation}\label{eq:bc_tang_boundarylayer}
    \lb  \delta v_y''(y=0)=\delta v_y'(y=0).
\end{equation}
Since $W$ is not present in this problem we use in this section $D$ as unit of length and $v_0$ as unit of velocity.
Only in this section we will define $\y = y/D$, $\tilde{k} = kD$, $\lbr = \lb/D$, $\reyi = v_0D/nu$, $\tilde{\omega} = \omega D/v_0$ and $\tilde{\tau} = \tau v_0/D$. 

A set of basis function adapted to this geometry and respecting the boundary conditions \eqref{eq:bc_normal_boundarylayer}-\eqref{eq:bc_tang_boundarylayer} is given by
\begin{equation}
    f_n(\y)=[L_n(\y) +c_0^{(n)} L_0(\y) +c_1^{(n)} L_1(\y)]e^{-\y/2},
\end{equation}
where $L_n(x)$ is the Laguerre polynomial of order $n$ \cite{noauthor_nist_2026} and the constants $c_i^{(n)}$ are chosen to enforce the boundary conditions
\begin{equation}
    f_n(0)=0,
\end{equation}
\begin{equation}
    \lbr f_n''(0)=f_n'(0),
\end{equation}
leading to
\begin{align}
    c_0^{(n)} &= \frac{n-1+\lbr(n^2/2+n/2-1)}{1+\lbr},\\
    c_1^{(n)} &= -\frac{n+\lbr(n^2/2+n/2)}{1+\lbr}.
\end{align}
Note that the functions $f_0$ and $f_1$ vanish identically and must be excluded from the expansion of $\delta v_y$.

By repeating the same steps as in Appendix~\ref{app:eigenvalues} we obtain an eigenvalue problem in the same form \eqref{eq:eigenvalue_problem} with the matrices $L$ and $B$ given by \eqref{eq:L_definition} and \eqref{eq:B_definition} but with coefficients given by
\begin{align}\label{eq:vy_coefficients_boundary}
a_4  & = -\frac{i}{\reyi}, \\
a_2 & = \tilde{k}\left(1+\frac{2i\tilde{k}}{\reyi} \right),\\
b_2 & = - \frac{\tilde{k}}{1  +\lbr},\\
a_0 & = -\tilde{k}^3\left(1+i\frac{\tilde{k}}{\reyi} \right),\\
b_0 & = \frac{\tilde{k}\left(\tilde{k}^2+1\right)}{1 +\lbr },
\end{align}
and matrices given by
\begin{align}\label{eq:laguerre_matrix_elements}
    M^{(0)}_{mn}&= \int_0^\infty f_{m}(\y)f_{n}(\y)d\y,\\
    M^{(2)}_{mn}&=\int_0^\infty f_{m}(\y)f_{n}''(\y)d\y,\\
    M^{(4)}_{mn}&= \int_0^\infty f_{m}(\y)f_{n}''''(\y)d\y,\\
    N^{(0)}_{mn}&= \int_0^\infty f_{m}(\y)e^{-\y}f_{n}(\y)d\y,\\  
    N^{(2)}_{mn}&= \int_0^\infty f_{m}(\y)e^{-\y}f_{n}''(\y)d\y. 
\end{align}
Explicit expressions of the matrix elements \eqref{eq:laguerre_matrix_elements} can be found in \cite{noauthor_iacopo_2026} together with Python functions implementing them.
Identifying the eigenvalues $\lambda_n$ of \eqref{eq:eigenvalue_problem} with $\tilde{\omega_n} +i\tilde{\tau}^{-1}$ yields this time
\begin{equation}
    \omega_n = \frac{v_0}{D}\left(\lambda_n -\frac{i}{\reyi}\right).
\end{equation}
\end{document}